\documentclass[reprint,amsmath,amssymb, aps, pra, showkeys,showpacs]{revtex4-2}
\usepackage{amsmath,amssymb, physics, braket, mathtools, graphicx,epstopdf,lastpage,hyperref}
\hypersetup{colorlinks=true, urlcolor=blue, citecolor=blue, linkcolor=blue}
\allowdisplaybreaks
\begin{document}
\title{Analyzing performance of $f$-deformed displaced Fock state in continuous-variable quantum teleportation}
\author{Deepak}
\email{deepak20dalal19@gmail.com}

\author{Arpita Chatterjee}
\email{arpita.sps@gmail.com}
\affiliation{Department of Mathematics\\J. C. Bose University of Science and Technology, YMCA\\Faridabad 121006, Haryana, 
India}
\date{\today}

\begin{abstract}
Performing non-Gaussian operations, namely photon addition, photon subtraction, photon-addition-then-subtraction, photon-subtraction-then-addition can successfully enhance the fidelity of the continuous-variable quantum teleportation. However, a shortcoming of these non-Gaussian resources is that they are probabilistic in nature. In this article, we investigate the success probability of the non-Gaussian resources for optimal performance of the ideal teleportation protocol. To this end, we first derive the analytical expression for the two-mode entangled channel having $f$-deformed displaced Fock state or photon-added displaced Fock state or photon-subtracted displaced Fock state at one port and vacuum at another port of a symmetric beam-splitter. The generalized displaced Fock states are obtained by replacing the conventional bosonic functions by the nonlinear $f$-deformed operators such as $A=af(a^\dag a)$ and $B=af(a^\dag a)^{-1}$. The Wigner characteristic functions describing these three aforementioned non-Gaussian states are determined and utilized to attain the teleportation fidelity for input coherent and squeezed vacuum states. It is found that the nonlinear substitution leads to an enhancement in teleportation fidelity beyond the threshold limit. Moreover, the entangled photon-subtracted displaced Fock state demonstrates maximum efficiency as a quantum channel for teleporting single-mode coherent and squeezed states. We provide the squeezing regime values corresponding to the optimal performance of the non-Gaussian states considered, which will be of significant interest to the experimental fraternity. Further, we show that the entangled photon-added displaced Fock states have larger amount of entanglement but the entangled photon-subtracted displaced Fock states have stronger Einstein-Podolsky-Rosen correlation.
\end{abstract}



\maketitle

\section{Introduction}
\label{int}

It is well known that Gaussian type entangled resource such as a two-mode squeezed vacuum state (TMSV) and Gaussian operations implemented by any combination of linear, quadratic optics, and
homodyne detection are generally employed in continuous-variable quantum information processing tasks \cite{Malpani2021,Deepak2023}. However, it has been realized that both Gaussian operations and entangled resources have some limitations. For example, quantum speed up for harmonic oscillators cannot be attained only by Gaussian operations with Gaussian inputs \cite{Bartlett2002,Bartlett2002a}, entanglement distillation from two Gaussian entangled states is not possible only by Gaussian local operations and classical communication \cite{Eisert2002,Fiurasek2002,Giedke2002}. Moreover, it has been proved recently that using Gaussian operations in quantum communication protocols cannot protect Gaussian states from Gaussian errors \cite{Niset2009}. Therefore, it becomes essential to look for non-Gaussian resources and operations which are more efficient in quantum information processing. One possible method for producing non-Gaussian entangled resources is to perform photon addition and/or subtractio on a given Gaussian state \cite{Malpani2021,Deepak2023}. Current advancements in experimental techniques have enabled the performance of photon addition and/or subtraction on a given input state. 

In recent time, Priya et. al. \cite{Malpani2019} studied the nonclassical properties of a set of engineered quantum states (both photon added and photon subtracted) using the lower and higher-order nonclassicality criteria. In some other work, they have quantified the nonclassicality and non-Gaussianity of photon-added displaced Fock states and discussed their significance in quantum technology and the necessity to quantify these resources \cite{Malpani2021}. An entangled two-mode state can be originated by employing a 50:50 symmetric beam-splitter (BS) while a Fock state and a vacuum are injected at the input modes of the BS. Consequently, inserting a class of generalized Fock states, e.g. displaced Fock state (DFS), photon-added displaced Fock state (PADFS), and photon-subtracted displaced Fock state (PSDFS) etc. from one input port and vacuum from the other of a 50:50 BS can manufacture two-mode entangled DFS (TMEDFS), two-mode entangled PADFS (TMEPADFS), and two-mode entangled PSDFS (TMEPSDFS) respectively. Deepak et al. \cite{Deepak2023} performed a quantitative investigation of the       
nonclassical and quantum non-Gaussian characters of the photon-subtracted displaced Fock stste by using a collection of measures like Wigner logarithmic negativity, linear entropy potential, skew information-based measure, and relative entropy of quantum non-Gaussianity. In present work, the performance of the TMEPADFS and TMEPSDFS as a quantum channel for continuous-variable quantum teleportation is investigated. In this approach, the input state is assumed to be either coherent or squeezed, and the success under the Braunstein and Kimble protocol is evaluated in terms of the average fidelity for transporting that single-mode optical state. At the same time, we work to comprehend how various realistic parameters contribute to successful teleportation of a single-mode input state. The advantage of $f$-deformed entangled resources in continuous-variable treleportation protocol under ideal circumstances is also explored. 

The idea of quantum teleportation was first proposed by Bennett et al. in the context of discrete variables \cite{Bennett1993}, and later illustrated experimentally by Bouwmeester \cite{Bouwmeester1997} and Popescu \cite{Boschi1998}. Vaidman introduced the concept of quantum teleportation in the continuous-variable regime \cite{Vaidman1994}. The actual quantum-optical protocol for transmitting quadrature amplitudes of a light field from a sender to a receiver in a distant place was developed by Braunstein and Kimble \cite{Braunstein1998} and implemented experimentally by Furusawa et al. \cite{Bowen2003,Furusawa1998,Takei2005} soon afterward. In a typical teleportation protocol, two users, Alice and Bob, shared an entangled resource of two modes $A$ and $B$. The state to be teleported is a single-mode input state, called $\ket{\text{in}}$ in the sender's place. The input mode $\ket{\text{in}}$ and the mode $A$ of the entangled channel are mixed at a 50:50 beam-splitter yielding the output modes $\ket{\text{out}}$ and $A'$. Then Alice performed a destructive homodyne measurement on the output modes and classically communicated the result to Bob. Bob applied a unitary displacement operation on mode $B$ to receive the desired teleported state. Different alternative descriptions of the original Braunstein-Kimble protocol have been used in the literature \cite{Enk1999,Vukics2002,Hofmann2000}. The quantum information fraternity is particularly interested in usage of non-Gaussian resources for improving the theoretical description of CV quantum teleportation and its experimental implementation  due to the escalating application of CV entangled resources in quantum information processing and quantum computing \cite{NavarreteBenlloch2015}. Very recently, Deepak et al. \cite{Deepak2022} described the ideal and realistic continuous-variable quantum teleportation protocol in terms of the characteristic functions of the quantum states involved (input, resource, and teleported states). Here we have extended that work by employing the generalized two-mode $f$-deformed entangled resources as quantum channels.

Entanglement and Einstein-Podolsky-Rosen (EPR) correlation are fundamental phenomena in the context of quantum mechanics. These concepts probe into the mysterious and paradoxical nature of the quantum world, challenging our classical understanding of reality. Entanglement refers to a unique quantum phenomenon where two or more particles become interconnected in such a way that the state of one particle instantaneously influences the state of another, irrespective of the physical distance between them. This correlation exists even when the particles are in remote location, defying our classical notions of causality and locality. EPR correlation, proposed by Einstein, Podolsky and Rosen, is an important aspect of entanglement. In their pioneering research \cite{Rosen1979}, these physicists proposed a hypothetical experiment to illustrate what they have noticed as a paradox within quantum mechanics. They have found that two quantum-entangled particles can have perfectly correlated positions and momenta that means the measurement of one entangled particle's property could immediately determine the state of another, that violates the principle of local realism where nothing can travel faster than the speed of light. In this paper, the entanglement amount and the EPR correlation of the non-Gaussian states of interest are investigated. These states are obtained by performing the nonlinear $f$-deformation and mixing a photon-added/subtracted displaced Fock state with a vacuum at a 50:50 beam-splitter. The performance of these non-Gaussian entangled resources in continuous-variable quantum teleportation of Braunstein and Kimble protocol for coherent states, and squeezed states is studied. 

The paper is structured as follows: the characteristic functions of all quantum states under consideration are studied in Sec.~\ref{qsi}. In Sec.~\ref{faeprc}, we have calculated the fidelity for continuous-variable quantum teleportation based on the non-Gaussian entangled resources. In Sec.~\ref{entepr}, the entanglement and EPR correlation are obtained. In the last Sec.~\ref{res}, we have summarized the main results. 

\section{$f$-deformed Displaced Fock States}
\label{qsi}

A displaced Fock state (DFS) described as a generalized coherent state, is defined by $\ket{\psi}=D(\alpha)\ket{n}$ where $D(\alpha)$ is the displacement operator and $\ket{n}$ is an arbitrary Fock state. In special case ($n=0$), this state reduces to a displaced vacuum state, also known as coherent state. If $\alpha=0$, the displaced Fock state represents a number state.

The wave functions for coherent, squeezed, $k$-photon-added and $k$-photon-subtracted displaced Fock states may be described as
\begin{align}
\label{wfqsi}
\ket{\psi_{\textrm{coh}}}& = \ket{\alpha_0}=D(\alpha_0)\ket{0}\nonumber\\
\ket{\psi_{\textrm{squ}}} &= \ket{z} = S(z)\ket{0},\,\,z=r e^{i\zeta}\\
\ket{\psi_+}& = N_+a^{\dag k}D(\alpha)\ket{n}\nonumber\\
\ket{\psi_-}& = N_-a^kD(\alpha)\ket{n}\nonumber
\end{align}
Here subscripts $\pm$ correspond to photon addition (photon subtraction), $N^{-2}_+=\sum_{p=0}^k{k\choose p}^2|\alpha|^{2(k-p)}\frac{(n+p)!}{n!}$ and $N^{-2}_-=\sum_{p=0}^k{k\choose p}^2|\alpha|^{2(k-p)}\frac{n!}{(n-p)!}$ are the normalization constants for $k$-photon-added displaced Fock state (PADFS) and $k$-photon-subtracted displaced Fock state (PSDFS), respectively. These states can be expressed in terms of the Fock state basis as following:
\begin{equation}
\ket{\psi_\pm}= \sum_{m=0}^\infty C_m^\pm(n,k,\alpha)\ket{m}
\end{equation}
where $C_m^+(n,k,\alpha)= N_+\bra{m}a^{\dag k} D(\alpha)\ket{n}$ and $C_m^-(n,k,\alpha)= N_-\bra{m}a^k D(\alpha)\ket{n}$. To investigate the performance of two classes of non-Gaussian entangled resources for teleportation of input coherent and squeezed states, we require the two-mode entangled channel $\rho_{AB}$ that is originated by mixing a PADFS (PSDFS) and a vacuum at a 50:50 beam-splitter (BS). The symmetric beam-splitter performs a balanced mixing of a Fock state and a vacuum injected at its input ports through a $SU(2)$ transformation that yields \cite{Deepak2023}:
\begin{equation}
\label{bse}
\ket{n}\otimes \ket{0} =\ket{n,0} \xmapsto{\textbf{BS}} \frac{1}{2^{n/2}}\sum_{j=0}^n\sqrt{{n\choose j}}\ket{j,n-j}
\end{equation}
Thus inserting PADFS (PSDFS) and vacuum at the input modes of the 50:50 beam-splitter, the post-BS two-mode entangled photon-added displaced Fock state (TMEPADFS) (two-mode entangled photon-subtracted displaced Fock state (TMEPSDFS)) $\rho_{AB}=\ket{\phi_\pm}\bra{\phi_\pm}$ can be obtained as 
\begin{align}\nonumber 
\label{kpstm}
\ket{\phi_\pm} &=  \ket{\psi_\pm}\otimes \ket{0}\\
& \xmapsto{\textbf{BS}} \sum_{m=0}^\infty \frac{C_m^\pm(n,k,\alpha)}{2^{m/2}}\sum_{j=0}^m\sqrt{m\choose j}\ket{j,m-j}
\end{align}
In order to introduce nonlinearity to TMEPADFS (TMEPSDFS), the generalized displacement operator $D'_f(\alpha)$ may be of use. This can be accomplished by recalling the first postulate of the nonlinear coherent states approach by Man'ko et. al. \cite{Man’ko1997}. The postulate affirms that the standard bosonic operators undergo $f$-deformation by applying an intensity-dependent function as follows:
\begin{align}\nonumber
\label{fdo}
A & =af(N)=f(N+1)a,\\\nonumber
A^\dag & =f^\dag(N)a^\dag=a^\dag f^\dag(N+1)
\end{align}
Here $f(N)$ is a number operator-valued ($N\equiv a^\dag a$)  nonlinear function of the intensity of light. Considering the $f$-deformed (nonlinear) bosonic operators, the generalized displacement operator $D_f(\alpha)$ can be written as
\begin{equation}
\label{fddo}
D_f(\alpha)=\exp(\alpha A^\dag-\alpha^* A)
\end{equation}
For further expansion of this generalized displacement operator, it is necessary to find a way to disentangle the $f$-deformed bosonic operators using Baker-Campbell-Hausdorf formula. The method of auxiliary operators suggests two auxiliary operators, namely $B = a{f(N)}^{-1}$ and $B^\dag = {f^{\dag}(N)}^{-1} a^\dag$ to overcome this difficulty \cite{Faghihi2020}. This leads to the generators $\{A, B^\dag, B^\dag A, I\}$ and $\{B, A^\dag, A^\dag B, I\}$ that satisfy the Weyl–Heisenberg Lie
algebra as well as the commutation relations $[A,~B^\dag A]=A$, $[B^\dag,~B^\dag A]=-B^\dag$, and $ [A,~B^\dag]=[B,~A^\dag]=I$. As a consequence of the auxiliary operators method, two generalized displacement operators are defined as
\begin{align}
D'_f&=\exp(\alpha A^\dag-\alpha^*B)\nonumber\\ &=\exp(-|\alpha|^2/2)\exp(\alpha A^\dag)\exp(-\alpha^* B)\\
D''_f&=\exp(\alpha B^\dag-\alpha^*A)\nonumber\\ &=\exp(-|\alpha|^2/2)\exp(\alpha B^\dag)\exp(-\alpha^* A)
\end{align}
Considering the nonlinear displacement operator $D'_f(\alpha)$, the generalized $f$-deformed two-mode entangled displaced Fock resource states can be expressed as
\begin{equation}
\label{kpstm1}
\ket{\varphi_\pm}=\sum_{m=0}^\infty \frac{\mathbb{C}_m^\pm(n,k,\alpha)}{2^{m/2}}\sum_{j=0}^m\sqrt{m\choose j}\ket{j,m-j}
\end{equation}
where $\mathbb{C}_m^+(n,k,\alpha)= \mathbb{N}_+\bra{m}A^{\dag k} D'_f(\alpha)\ket{n}$ and $\mathbb{C}_m^-(n,k,\alpha)= \mathbb{N}_-\bra{m}A^k D'_f(\alpha)\ket{n}$, $\mathbb{N}_\pm$ are the corresponding normalization constants.

In the continuous-variable teloportation following the Braunstein-Kimble (BK) protocol, the fidelity is described in terms of the characteristic functions of the quantum states involved (input, resource, and teleported states) \cite{Marian2006}. This formalism is especially beneficial while dealing with
non-Gaussian states and resources because it very much simplifies calculation complexities. The characteristic functions for a coherent state $\ket{\alpha_0}$, a squeezed state $\ket{z}$, and a TMEPADFS (TMEPSDFS) $\ket{\phi_\pm}$ can be obtained as
\begin{align}
\label{cfqsi}
\chi_{\textrm{coh}}(\gamma) & =\exp(\frac{-|\gamma|^2}{2}+\alpha_0^\star\gamma-\alpha_0\gamma^\star),
\end{align}
\begin{align}
\chi_{\textrm{squ}}(\gamma) &=\exp(-\frac{2|\gamma|^2\cosh {2r}+(\gamma^2e^{-\iota\zeta}+\gamma^{\star 2}e^{\iota\zeta})\sinh{2r}}{4}) 
\end{align}
and
\begin{align}
\chi_{12}^\pm(\gamma_1,\gamma_2) &= \exp(-|\gamma_1|^2/2-|\gamma_2|^2/2)\nonumber \\ & \times \sum_{m,l=0}^\infty \frac{\left(C_m^\pm(n,k,\alpha)\right)^*C_l^\pm (n,k,\alpha)}{2^{(m+l)/2}}\sqrt{\frac{l!}{m!}}\gamma_2^{m-l}\nonumber \\ & \times \sum_{j,k=0}^{m,l}{m\choose j} \gamma_1^{j-k}\gamma_2^{k-j}L_k^{j-k}(|\gamma_1|^2)L_{l-k}^{m-l+k-j}(|\gamma_2|^2)
\end{align}
where $L_n^k(x)$ is the associated Laguerre polynomial of order $n$.
\section{Teleportation using $f$-deformed two-mode entangled resources}
\label{faeprc}

The idea of quantum teleportation was first proposed by Bennett et al. in the discrete variable regime \cite{Bennett1993}. Later on, Braunstein and Kimble (BK) introduced the actual continuous-variable (CV) quantum-optical protocol for teleporting the quadrature amplitudes of a light field that employed the Wigner functions of the associated quantum states \cite{Braunstein1998}. 
For any given state, there is a one-to-one correspondence between its Wigner function and symmetrically ordered characteristic function. The BK protocol in terms of the characteristic functions is particularly suitable while employing non-Gaussian states as it greatly simplifies the calculation complexities. 

In the standard BK protocol of continuous-variables, two users, Alice and Bob shared a two-mode entangled state of modes $A$ and $B$. A single-mode input quantum state to be teleported is provided to Alice. The characteristic functions of the single-mode input state to be teleported and the two-mode entangled resource state are denoted by $\chi_{\textrm{in}}(\gamma)$ and $\chi_{\textrm{res}}(\gamma^*,\gamma)$, respectively. The input state and mode $A$ of the entangled resource are mixed at a 50:50 beam-splitter. After that, the sender performed a destructive homodyne measurement at the output modes of the BS and the results are classically communicated to the receiver. Based on Alice's outcome, Bob applied a specific set of quantum gates on mode $B$ of the entangled pair and displayed his result leading to the teleported state. By exploiting the Weyl expansion, the BK protocol allows one to write the characteristic function of the teleported state in a factorized form as
\begin{equation}
\chi_{\textrm{out}}(\gamma) = \chi_{\textrm{in}}(\gamma)\chi_{\textrm{res}}(\gamma^*,\gamma)
\end{equation}
In order to measure the probability of success in a teleportation protocol, it is convenient to use the teleportation fidelity $F$. This is a state-dependent quantity that measures the overlap between the input state $\rho_{\textrm{in}}$ and the output (teleported) state $\rho_{\textrm{out}}$, i.e., $F=\textrm{Tr}[\rho_{\textrm{in}}\rho_{\textrm{out}}]$.  If the input and output states are identical, then $F=1$ corresponding that perfect teleportation has been reached \cite{Quintero2011}. The fidelity in terms of the associated characteristic functions reads as \cite{Dell’Anno2007,Dell’Anno2010}
\begin{equation}
\label{fdef}
F=\frac{1}{\pi}\int d^2\gamma \chi_{\textrm{in}}(\gamma)\chi_{\textrm{out}}(-\gamma),
\end{equation}
It has been found that a maximum fidelity of 1/2 can be achieved when an input coherent state is teleported  without using a shared entangled state. Therefore, a successful quantum
teleportation is observed when the fidelity value surpasses the classical limit of 1/2. We require an infinitely entangled resource state to achieve perfect teleportation with unit
fidelity. Further, the fidelity should be greater than 2/3 for secure
teleportation, i.e., to ensure that any copy of the teleported state is not possible by an eavesdropper \cite{chandan}. 

In the following, we will make use of \eqref{fdef} to derive the fidelity for teleportation of input coherent and squeezed vacuum states and to quantify the success of the protocol for different classes of non-Gaussian entangled resources.

\subsection{Input Coherent State}

The fidelity for teleporting an input coherent state via TMEPADFS (TMEPSDFS) resource states can be evaluated using \eqref{fdef} as
\begin{align}
\label{f1eq}
F({\phi_\pm^{\textrm{coh}}})&=\sum_{m,l=0}^\infty \frac{(C_m^\pm(n,k,\alpha))^*C_l^\pm(n,k,\alpha)}{2^{m+1}}\sqrt{\frac{l!}{m!}}(-1)^{m-l} \nonumber\\ & \times \sum_{j=0}^{m}{m\choose j}\sum_{r,s=0}^{j+\frac{l-m}{2},-j+\frac{l+m}{2}} {j\choose j+\frac{l-m}{2}-r} \\ & \times {m-j\choose -j+\frac{l+m}{2}-s}\frac{(-1)^{r+s}}{2^{r+s}r!s!}\left(\frac{m-l}{2}+r+s\right)!\nonumber 
\end{align}

\begin{figure*}[hbt]
\centering
\includegraphics[scale=1.25]{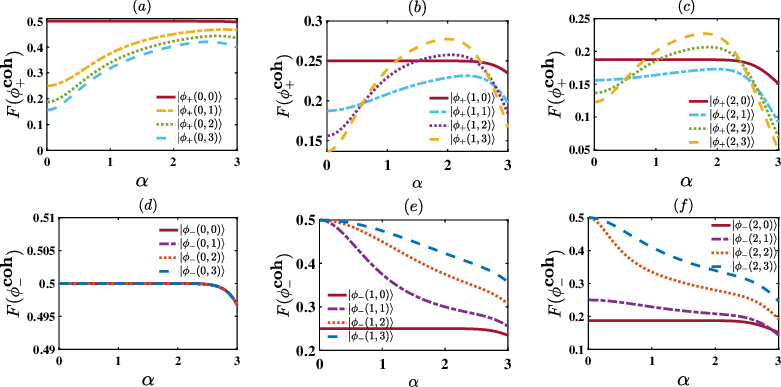}
\caption{(Color online) Variation of fidelity $F({\phi_\pm^{\textrm{coh}}})$ with respect to TMEPADFS (TMEPSDFS) parameter $\alpha$ and for different values of $n$ and $k$ with $f(n)=1$.}
\label{figcs}
\end{figure*}

In order to find the teleportation fidelity using the $f$-deformed two-mode entangled resources, coefficients $C_m^\pm(n,k,\alpha)$ in \eqref{f1eq} will be replaced by $\mathbb{C}_m^\pm(n,k,\alpha)$. 
It can be seen that the fidelity depends only on the Fock parameter $n$, photon-addition/subtraction number $k$, coherent state amplitude $\alpha$ of the entangled resources and is free from the amplitude $\alpha_0$ of the input coherent state.

The fidelity for teleporting a coherent state is plotted as a function of amplitude $\alpha$ of entangled resources in Fig.~\ref{figcs}. It is noted that for $n=k=0$, the TMEPADFS switches to a two-mode entangled coherent state and the corresponding fidelity approaches the maximum threshold limit of 0.5 (see Fig.~\ref{figcs}(a)). The fidelity with the TMEPADFS is less than that with the original no photon-added DFS. It is also observed that TMEPADFS yields lower fidelity value compared to TMEPSDFS although PADFS is more nonclassical than PSDFS \cite{Malpani2021,Malpani2019}. Moreover,  the fidelity plot goes up with an increase in the photon subtraction number but diminishes as the photon addition number is increasing. For a fixed value of photon-addition/subtraction number $k$, an increase in the Fock parameter $n$ leads to weak fidelity, while varying the displacement parameter $\alpha$ from 0 to 3 increases the fidelity value in the beginning followed by a collapse. In case of $n=0$, PSDFS reduces to the elementary coherent state, causing fidelity to coincide with that of DFS for different values of the photon-subtraction number $k$ (see Fig.~\ref{figcs}(d)). We see that for the linear intensity-dependent function $f(n)=1$, the fidelity remains below the classical limit of 1/2, thus fails to prove linear TMEPADFS and TMEPSDFS relevant for quantum teleportation. 

\begin{figure*}[hbt]
\includegraphics[scale=1.25]{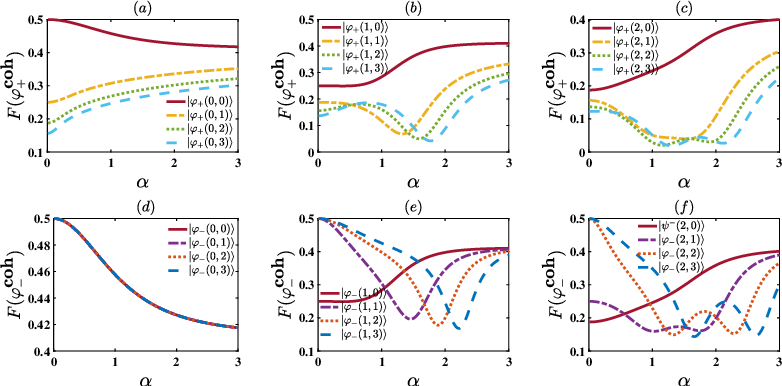}
\caption{(Color online) Behaviour of fidelity $F({\varphi_\pm^{\textrm{coh}}})$ associated to $f$-deformed entangled resource states, plotted with respect to $\alpha$ and for different values of $n$ and $k$ with $f(n)=\sqrt{n}$.}
\label{fig5}
\end{figure*}

\begin{figure*}[hbt]
\includegraphics[scale=1.25]{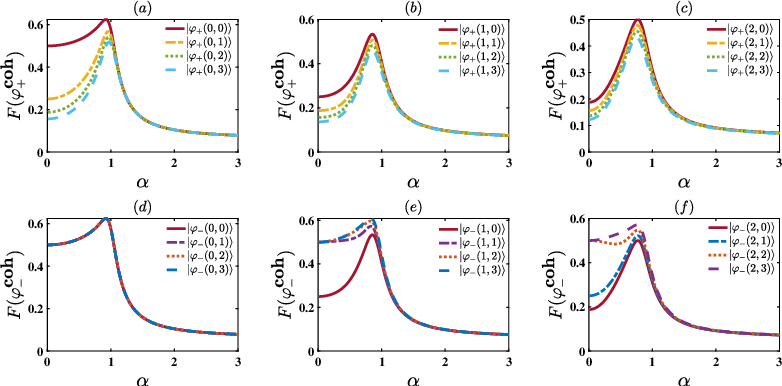}
\caption{(Color online) Plot of fidelity $F({\varphi_\pm^{\textrm{coh}}})$ as a function of $\alpha$ for different values of $n$, $k$ and $f(n)=1/\sqrt{n}$.}
\label{fig3}
\end{figure*}

To overcome this challenge, we consider next the teleportation fidelity of a coherent state with different $f$-deformed non-Gaussian entangled resources. The fidelities corresponding to the nonlinear intensity-dependent functions $f(n)=\sqrt{n}$ and $f(n)=1/\sqrt{n}$ are plotted in Figs.~\ref{fig5} and \ref{fig3}, respectively. Comparing with panels I and II of Fig.~\ref{figcs}, we can see that the qualitative behaviours are very similar to the previous cases of linear TMEPADFS and TMEPSDFS. 
In Fig.~\ref{fig5}, the fidelity initially experiences a decline and then ascends with the non-Gaussian state channel parameter $\alpha$. However, it also remains unable to cross the classical boundary. As the Fock parameter or photon addition number increases, fidelity decreases while an increase in the photon subtraction parameter results enhancement in fidelity. Fig.~\ref{fig3} shows a different fidelity pattern for the intensity-dependent function $f(n)=1/\sqrt{n}$. In this case, the fidelity plot outperforms the threshold limit that leads to a significant advantage of using $f$-deformed resource states. Similar to Fig.~\ref{fig5}, an increase in the Fock parameter or photon addition number corresponds to a fidelity reduction, while an increase in the photon subtraction parameter leads to improvement of fidelity.

\subsection{Input Squeezed State}

\begin{figure*}[hbt]
\centering
\includegraphics[scale=1.25]{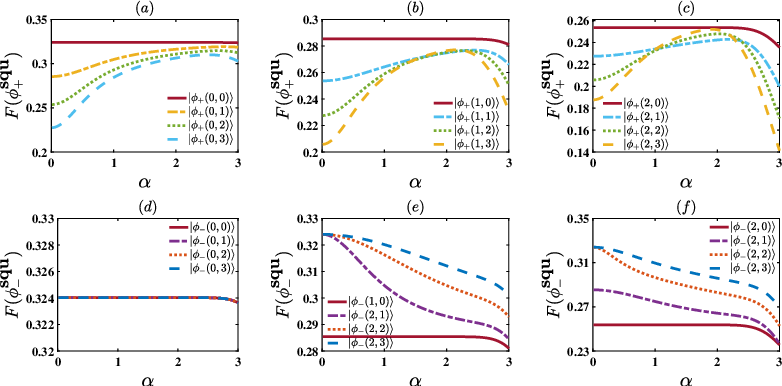}
\caption{(Color online) Fidelity of teleportation $F({\phi_\pm^{\textrm{squ}}})$ as a function of $\alpha$ with $r=1$ and $f(n)=1$.}
\label{figcs1}
\end{figure*}

\begin{figure*}[hbt]
\centering
\includegraphics[scale=1.25]{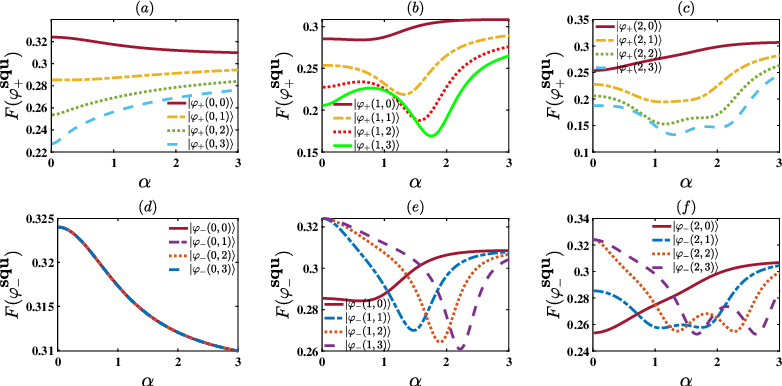}
\caption{(Color online) Fidelity $F({\varphi_\pm^{\textrm{squ}}})$ is plotted with respect to $\alpha$ with $r=1$ and $f(n)=\sqrt{n}$.}
\label{fig6}
\end{figure*}

\begin{figure*}[hbt]
\centering
\includegraphics[scale=1.25]{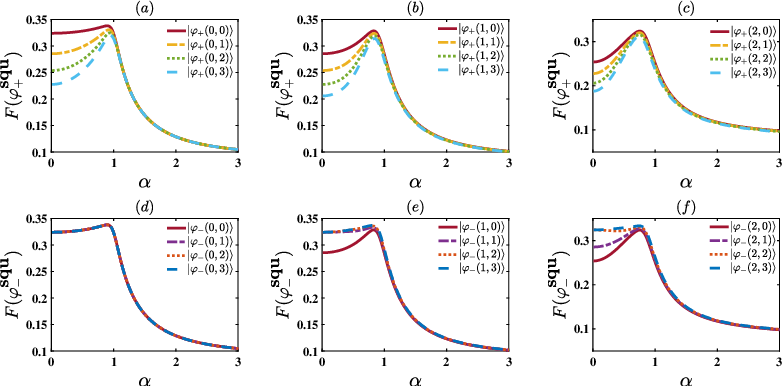}
\caption{(Color online) Variation of fidelity $F({\varphi_\pm^{\textrm{squ}}})$ with respect to $\alpha$ with $r=1$ and $f(n)=1/\sqrt{n}$.}
\label{fig4}
\end{figure*}

Once the characteristic functions of the two-mode entangled photon-added/subtracted displaced Fock states are derived, we now proceed to calculate the fidelity for teleporting an input squeezed vacuum state $\ket{z}$ with $z=r\,e^{i\zeta}$. The BK protocol is followed for teleporting an input quantum state between two distant physical systems. The analytic expression of fidelity using TMEPADFS (TMEPSDFS) channel turns out to be
\begin{align}
\label{f2eq}
& 
F({\phi_\pm^{\textrm{squ}}})\nonumber \\ &= \frac{1}{2\cosh{r}}\sum_{m,l=0}^\infty \frac{(C_m^\pm(n,k,\alpha))^*C_l(n,k,\alpha)}{2^{(m+l)/2}}\sqrt{\frac{l!}{m!}}\nonumber \\ & \times (-1)^{m-l}\sum_{j,k=0}^{m,l}{m\choose j}\sum_{r,s=0}^{k,l-k} {j\choose k-r}{m-j\choose l-k-s}\frac{(-1)^{r+s}}{r!s!}\nonumber \\ & \times \sum_{t=0}^{j-k+r+s}\binom{j-k+r+s}{t}\frac{1}{2^t}\left(\frac{\tanh{r}}{2}\right)^{r+s-t+\frac{m-l}{2}}\nonumber \\ & \times \frac{(k-j+r+s+m-l)!}{(k-j+r+s+m-l-t)!!}\nonumber \\ & \times e^{i\zeta{(k-j+m-l)/2}}(j-k+r+s-t-1)!! 
\end{align}
if $k-j+r+s+m-l-t$ and $j-k+r+s-t$ are even and
\begin{eqnarray*}
		n!!=
		\left\{
		\begin{array}{lll}
			& n(n-2)(n-4)\ldots 4.2 & \mbox{if $n$ is even},\\\\
			& n(n-2)(n-4)\ldots3.1 & \mbox{if $n$ is odd}
		\end{array}
		\right.
	\end{eqnarray*}
$F({\phi_\pm^{\textrm{squ}}})$ are plotted as a function of $\alpha$ in Fig.~\ref{figcs1}.
We notice that in all the cases under consideration, $F({\phi_+^{\textrm{squ}}})$ initially increases until reaching a maximum limit and subsequently decreases. The fidelity pattern for an input squeezed state and TMEPADFS/TMEPSDFS/TMEDFS channels matches that of an input coherent state, although a bit lower. Notably, it equals to the fidelity of the input coherent state when the squeezing parameter $r$ is zero. To explore the fidelity for nonlinear intensity-dependent functions, we have plotted this for teleporting an input squeezed state via $f$-deformed channels such as TMEPADFS/TMEPSDFS/TMEDFS with $f(n)=\sqrt{n}$ and $f(n)=1/\sqrt{n}$ in Figs.~\ref{fig6} and \ref{fig4}, respectively.
Fig.~\ref{fig6} clearly shows that for the intensity-dependent function $f(n)=\sqrt{n}$, the teleportation fidelity initially decreases and then increases with the non-Gaussian state channel parameter $\alpha$. However, it remains unable to surpass the threshold limit. Increasing the Fock parameter or photon addition number results in fidelity reduction, while an increase in the photon subtraction number leads to fidelity improvement. Similarly in Fig.~\ref{fig4}, we observe that for the intensity-dependent function $f(n)=1/\sqrt{n}$, the teleportation fidelity first increases and then decreases with the non-Gaussian channel parameter $\alpha$. Like the previous case, it fails to cross the threshold limit. The fidelity decreases with an increase in the Fock parameter or photon addition number, while increases with an increase in the photon subtraction number.

\begin{figure*}[hbt]
\centering
\includegraphics[scale=1.25]{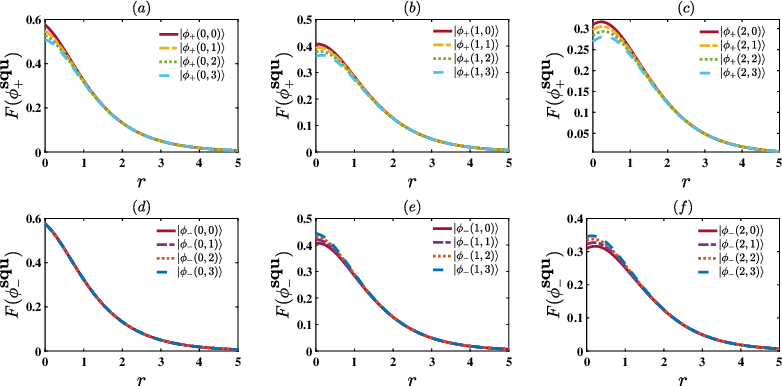}
\caption{(Color online) Fidelity $F({\phi_\pm^{\textrm{squ}}})$ as a function of squeezing parameter $z$ with $\alpha=1$ and $f(n)=1$.}
\label{fig7}
\end{figure*}

\begin{figure*}[hbt]
\centering
\includegraphics[scale=1.25]{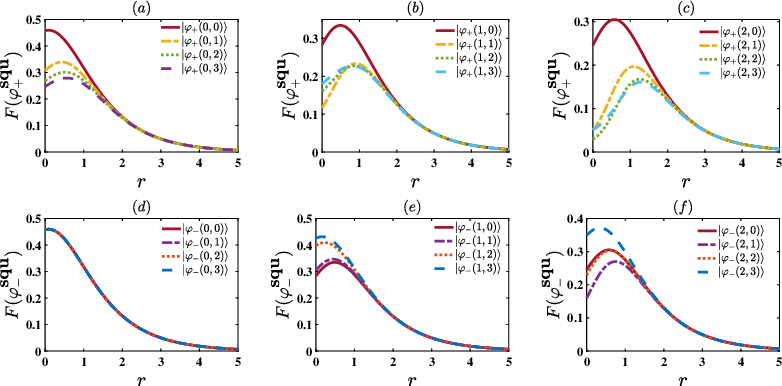}
\caption{(Color online) Fidelity $F({\varphi_\pm^{\textrm{squ}}})$ with respect to squeezing parameter $z$ of an input squeezed state $\ket{z}$ with $\alpha=1$ and $f(n)=\sqrt{n}$.}
\label{fig8}
\end{figure*}

\begin{figure*}[hbt]
\centering
\includegraphics[scale=1.25]{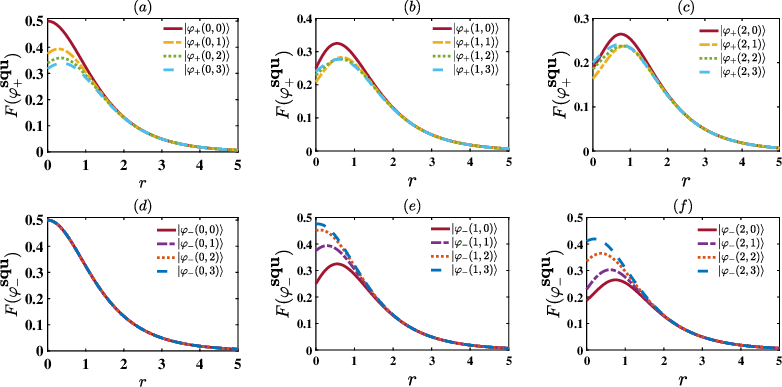}
\caption{(Color online) Variation of fidelity $F({\varphi_\pm^{\textrm{squ}}})$ as a function of $z$ with $\alpha=1$ and $f(n)=1/\sqrt{n}$.}
\label{fig9}
\end{figure*}

 To examine the performance of TMEPADFS/TMEPSDFS channels for teleporting an input squeezed vacuum state with respect to the squeezing parameter $z$, the variations of $F({\phi_\pm^{\textrm{squ}}})$ and $F({\varphi_\pm^{\textrm{squ}}})$ are plotted with respect to $z$ for $f(n)=1$, $\sqrt{n}$ and $1/\sqrt{n}$ in Figs.~\ref{fig7}, \ref{fig8} and \ref{fig9}, respectively.

These figures exhibit that the fidelity remains below the classical limit for different values of $f(n)$. Thus for squeezed state inputs, using nonlinear intensity-dependent functions has not been proved beneficial in terms of the fidelity. It is also evident that the fidelity reaches its maximum at a low squeezing regime and subsequently decreases. 

\section{Entanglement and EPR Correlation of $f$-deformed States}
\label{entepr}
\begin{figure*}
\includegraphics[scale=1.25]{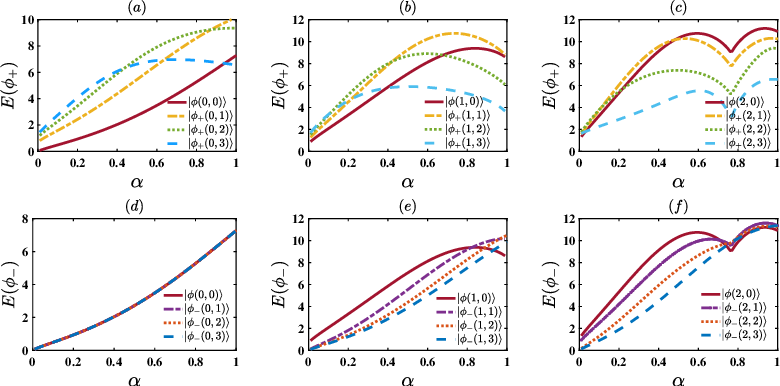}
\caption{(Color online) Entanglement of formation $E(\phi_\pm)$ as a function of TMEPADFS/TMEPSDFS parameter $\alpha$ for different values of $n$ and $k$, $f(n)=1$.}
\label{figent}
\end{figure*}

\begin{figure*}
\includegraphics[scale=1.25]{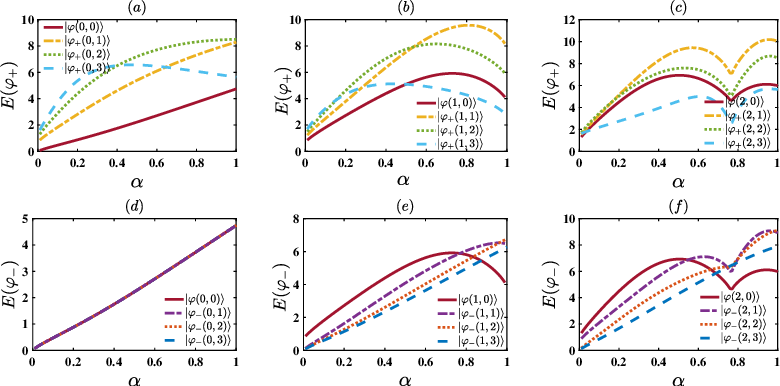}
\caption{(Color online) Entanglement of formation $E(\varphi_\pm)$ as a function of TMEPADFS/TMEPSDFS parameter $\alpha$ for different values of $n$ and $k$ and for $f(n)=\sqrt{n}$.}
\label{figentb}
\end{figure*}

\begin{figure*}
\includegraphics[scale=1.25]{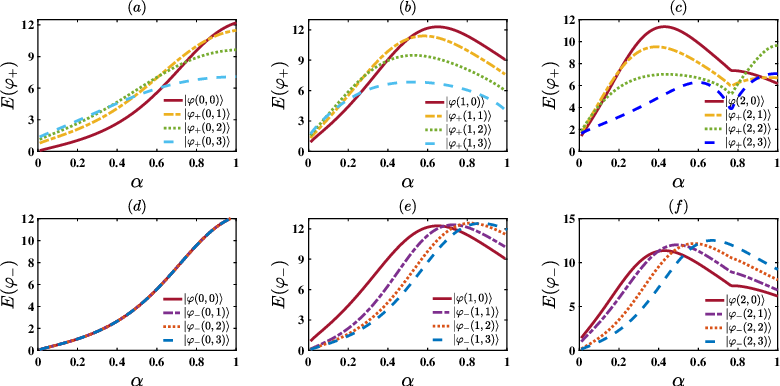}
\caption{(Color online) Entanglement of formation $E(\varphi_\pm)$ as a function of TMEPADFS/TMEPSDFS parameter $\alpha$ for different values of $n$ and $k$ and for $f(n)=1/\sqrt{n}$.}
\label{figentc}
\end{figure*}

Entangled resources are highly advantageous in various quantum information processing tasks such as continuous-variable quantum teleportation
\cite{Braunstein1998}, quantum dense coding \cite{PhysRevA.61.042302}, and entanglement swapping \cite{tan1999confirming}.  
Entanglement is a quantum mechanical phenomenon where two or more particles become correlated in such a way that the state of one particle cannot be described independently of the state of the other particle(s) \cite{Yang2009}. In other words, the particles become ``entangled" and their properties are reliant with each other. For any bipartite pure state, the entanglement can be quantified by using the partial von Neumann entropy as $E(\Psi_{12})=S(\rho_1)=-\Tr[\rho_1\ln(\rho_1)]$ i.e. entanglement of formation \cite{hbennet}. The entanglement for $f$-deformed TMEPADFS and TMEPSDFS can be calculated as
\begin{align}
\label{entdef2}
&E(\varphi_\pm)\nonumber\\ &=-\sum_{m,l,j=0}^{\infty,\infty,m}\frac{|\mathbb{C}_m^\pm(n,\alpha)\mathbb{C}_l^\pm(n,\alpha)|}{2^{(m+l)/2}}\sqrt{{m\choose j}{l\choose l-m+j}}\nonumber\\&\times\ln(\frac{|\mathbb{C}_m^\pm(n,\alpha)\mathbb{C}_l^\pm(n,\alpha)|}{2^{(m+l)/2}}\sqrt{{m\choose j}{l\choose l-m+j}})
\end{align}
The entanglement of these states is plotted in Figs.~\ref{figent}, \ref{figentb} and \ref{figentc}.

We observe that the entanglement of all the states under consideration first increases upto the maximum value and then decreases rapidly with respect to the displacement parameter $\alpha$. Further, increasing the photon addition enhances the entanglement while increasing the number of subtracted photons reduces the entanglement value. Also adding photons seems to be more significant than subtracting. Increasing the Fock state parameter $n$ increases the amount of entanglement.

In addition to the degree of entanglement, entangled states can also be characterized by the Einstein-Podolsky-Rosen (EPR) correlations between phase-quadrature components of the two modes. Analogous to the position and momentum operators of a macroscopic particle, the phase-quadrature operators of each mode are defined as $x_j=\frac{1}{\sqrt{2}}(a_j+a_j^\dag)$ and $p_j=\frac{1}{i\sqrt{2}}(a_j-a_j^\dag)$ for $j=1,2$. In the vacuum state, both the variances $\Delta(x_1-x_2)^2$ and $\Delta(p_1+p_2)^2$ are equal to one. For any two-mode classical state, these variances $\Delta(x_1-x_2)^2$ and $\Delta(p_1+p_2)^2$ are greater than 1. In the EPR state, $\Delta(x_1-x_2)^2=\Delta(p_1+p_2)^2=0$. That means $x_1$ and $p_1$ of 
the first mode can be exactly estimated by measured results of $x_2$ and $p_2$ of the second mode or vice versa. In this sense, the ideal EPR correlation exists between the two modes. There may be some two-mode states, e.g. two-mode squeezed state that possess the EPR correlation beyond the limit of the vacuum state, i.e., both the variances $\Delta(x_1-x_2)^2$ and $\Delta(p_1+p_2)^2$ are less than 1. This quantum correlation is a key ingredient for realizing quantum teleportation of continuous variables.

\begin{figure*}
\includegraphics[scale=1.25]{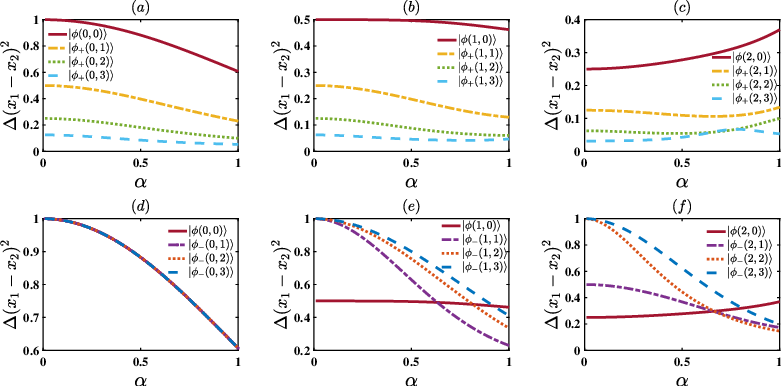}
\caption{(Color online) Variance $\Delta(x_1-x_2)^2$ for the state $\ket{\phi_\pm}$ with $f(n)=1$ and as a function of $\alpha$ for different values of $n$ and $k$.}
\label{figepr}
\end{figure*}

\begin{figure*}
\includegraphics[scale=1.25]{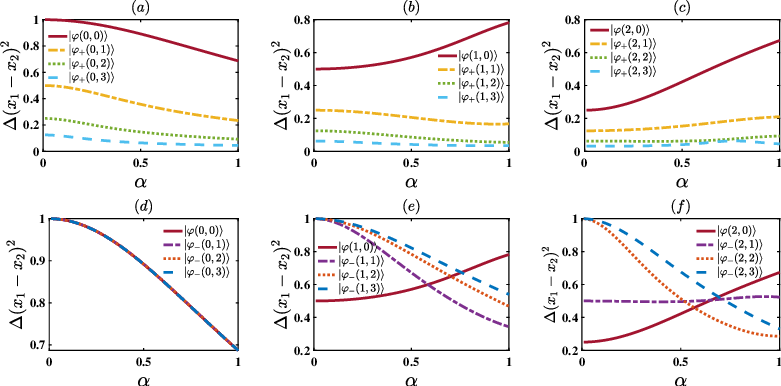}
\caption{(Color online) Variance $\Delta(x_1-x_2)^2$ for the state $\ket{\varphi_\pm}$ with $f(n)=\sqrt{n}$ and as a function of $\alpha$ for different values of $n$ and $k$.}
\label{figeprb}
\end{figure*}

\begin{figure*}
\includegraphics[scale=1.25]{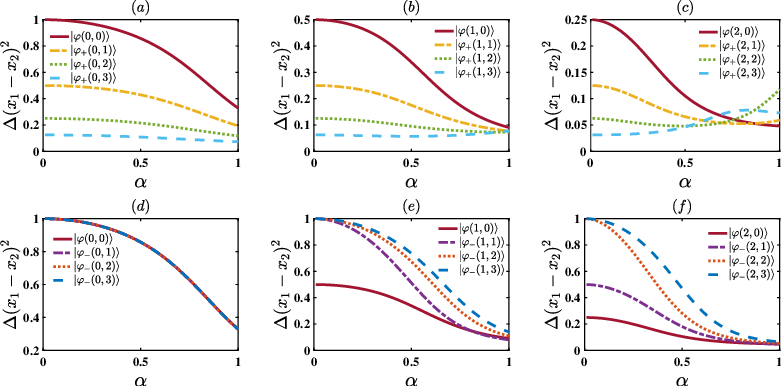}
\caption{(Color online) Variance $\Delta(x_1-x_2)^2$ for the state $\ket{\varphi_\pm}$ with $f(n)=1/\sqrt{n}$ and as a function of $\alpha$ for different values of $n$ and $k$.}
\label{figeprc}
\end{figure*}

For all the states, $\Delta(x_1-x_2)^2=\Delta(p_1+p_2)^2$. In Figs.~\ref{figepr}, \ref{figeprb} and \ref{figeprc}, the variance $\Delta(x_1-x_2)^2$ for TMEDFS, TMEPADFS, TMEPSDFS is plotted.  

It has been found that the variance $\Delta(x_1-x_2)^2$ decreases rapidly with respect to $\alpha$. The EPR correlation for TMEPADFS as well as TMEPSDFS decreases as $n$ changes from 0 to 2. We have further noticed that the variance of two-mode entangled photon-added displaced Fock state is less than that of two-mode entangled photon-subtracted displaced Fock state. Therefore, one may expect that the quality of quantum teleportation of continuous-variables can be
improved by using two-mode entangled photon-subtracted displaced Fock states.

\section{Conclusion}
\label{res}
In this work, the success probability for employing $f$-deformed two-mode entangled resources in continuous-variable quantum teleportation is evaluated. The entanglement properties of the non-Gaussian states are also investigated which are constructed by mixing a single-mode PADFS/PSDFS and a vacuum state at a 50:50 beam-splitter. We study quantum teleportation of Braunstein and Kimble
protocol for coherent states and squeezed states, with the resulting non-Gaussian states as entangled resources. For coherent states, the analytical expression of fidelity is derived. It is noted that the fidelity is independent of the amplitude of coherent states to be teleported.
For the squeezed vacuum input, the fidelity is state dependent. For all the states to be teleported, we have observed that the fidelity with the TMEPSDFS is higher than that with the TMEPADFS and it decreases as Fock parameter $n$ increases. The fidelity for teleporting coherent state with the non-Gaussian $f$-deformed two-mode entangled resources moves across the benchmark value of 1/2. Moreover, in the weak squeezing regime of the original squeezed state $\ket{z}$, the TMEPADFS and TMEPSDFS can lead to the highest
fidelity, although fail to cross the classical limit.

We have also observed that in a particular region,  the partial von Neumann entropy of all the resulting states is greater than that of the original two-mode entangled coherent state in a particular range of $\alpha$. The photon addition appears to be a more efficient operation for enhancing the amount of entanglement. As the EPR correlation between phase-quadrature components of the two modes in the bipartite states signals the
existence of entanglement, we have found that TMEPSDFS states have stronger EPR correlation than the TMEPADFS. Moreover, in the low-energy regime
of the input squeezed vacuum state, the plot for variance collapses quickly.
 
In addition, if the nonlinear intensity-dependent function can be fixed as $f(n)=1/n^p$, then for a large positive real number $p$, it becomes possible to achieve teleportation fidelity close to 1 for any of the input states.

\begin{center}
	\textbf{ACKNOWLEDGEMENT}
\end{center}
Deepak's work is supported by the Council of Scientific and Industrial Research (CSIR), Govt. of India (Award no. 09/1256(0006)/2019-EMR-1). A. C. acknowledges DST SERB for the support provided through the project number SUR/2022/000899.

\bibliography{pref_new.bib}
\end{document}